\documentclass[12pt]{article}
\usepackage{amsfonts,amsthm,amsmath,amssymb,graphicx,hyperref,color,yfonts}

\newcommand{\bea}{\begin{eqnarray}\displaystyle}
\newcommand{\eea}{\end{eqnarray}}
\newcommand{\KN}{\mathbin{\bigcirc\mspace{-15mu}\wedge\mspace{3mu}}}

\begin{document}
\makeatletter
\@addtoreset{equation}{section}
\makeatother
\renewcommand{\theequation}{\thesection.\arabic{equation}}
\vspace{1.8truecm}

{\LARGE{ \centerline{\bf Complexity from Spinning Primaries}}}  

\vskip.5cm 

\thispagestyle{empty} 
\centerline{ {\large\bf Robert de Mello Koch$^{a,b,}$\footnote{{\tt robert@neo.phys.wits.ac.za}},
Minkyoo Kim$^{b,c,}$\footnote{{\tt mimkim80@gmail.com}}  }}
\centerline{{\large\bf and
Hendrik J.R. Van Zyl${}^{b,}$\footnote{ {\tt hjrvanzyl@gmail.com}} }}

\vspace{.8cm}
\centerline{$^{a}${\it School of Science, Huzhou University, Huzhou 313000, China,}}

\vspace{.5cm}
\centerline{{\it ${}^{b}$ National Institute for Theoretical Physics,}}
\centerline{{\it School of Physics and Mandelstam Institute for Theoretical Physics,}}
\centerline{{\it University of the Witwatersrand, Wits, 2050, }}
\centerline{{\it South Africa }}

\vspace{.5cm}
\centerline{{\it $^{c}$ Center for Quantum Spacetime (CQUeST),}}
\centerline{{\it Sogang University Seoul, 121-742,}}
\centerline{{\it South Korea}}

\vspace{1truecm}

\thispagestyle{empty}

\centerline{\bf ABSTRACT}

\vskip.2cm 
We define circuits given by unitary representations of Lorentzian conformal field theory in 3 and 4 dimensions.
Our circuits start from a spinning primary state, allowing us to generalize formulas for the circuit complexity obtained
from circuits starting from scalar primary states.
These results are nicely reproduced in terms of the geometry of coadjoint orbits of the conformal group.
In contrast to the complexity geometry obtained from scalar primary states, the geometry is more complicated and the existence
of conjugate points, signaling the saturation of complexity, remains open.

\setcounter{page}{0}
\setcounter{tocdepth}{2}
\newpage
\tableofcontents
\setcounter{footnote}{0}
\linespread{1.1}
\parskip 4pt

{}~
{}~

\section{Introduction}

There is compelling evidence that quantum information has a role to play in the gauge theory / gravity 
duality\cite{Maldacena:1997re}.
A central concept in quantum information theory is entanglement between quantum states.
As a result of entanglement, even when the combined state of two particles can be completely specified, 
the state of each entangled particle can still be completely random when measured alone. 
Two-particle entanglement is well understood.
In stark contrast to this, there is no specific measure of the amount of entanglement for three or more particles. 
One useful measure of entanglement, the entanglement entropy, characterizes what a subsystem knows about the 
full quantum state.
Entanglement entropy marries the concept of entanglement with that of entropy, the degree of randomness of the system.
Motivated by gauge theory / gravity duality a connection between the entanglement entropy of a subregion ${\cal R}$ of the 
CFT, and the area of a minimal surface tethered to the boundary of ${\cal R}$ and exploring the bulk, is now 
known\cite{Ryu:2006bv,Ryu:2006ef,Hubeny:2007xt,Faulkner:2013ana,Engelhardt:2014gca}.
Developing this geometrical approach towards entanglement entropy, a remarkable identity between the connectivity 
of space and entanglement was described in \cite{VanRaamsdonk:2010pw}.
The geometrization of entanglement entropy has also provided deep insights into the information loss puzzle and into how
information is encoded in a quantum theory of 
gravity\cite{Penington:2019npb,Almheiri:2019psf,Penington:2019kki,Almheiri:2019qdq}.
This progress motivates entanglement entropy as an interesting quantity to study.
Entanglement entropy does not, however,  completely capture correlations in quantum states \cite{Susskind:2014moa}.
Another important measure of quantum information is quantum computational complexity \cite{Harlow:2013tf}.
Quantum computational complexity estimates how hard it is to construct a given quantum state, starting from
a fixed reference state and acting with elements of a set of elementary unitary operations\cite{Watrous,Aaronson}.
Computational complexity can be estimated by distances in the manifold of unitary operators and in this
way the description of complexity is naturally couched in the language of differential 
geometry\cite{Nielsen1,Nielsen2,Nielsen3}. 
Through the AdS/CFT correspondence there is another geometrical connection, which links complexity to the
geometry of black holes \cite{Susskind:2014moa,Susskind:2018pmk}.
Specifically, complexity is naturally related to the growth of the black hole interiors and the response of
complexity to perturbations resembles how the back hole interior reacts to 
perturbations\cite{Brown:2015lvg,Brown:2015bva,Carmi:2017jqz,Stanford:2014jda,Chapman:2018dem,Chapman:2018lsv}.
See also \cite{Brown:2017jil,Sun:2019yps,Bernamonti:2019zyy,Bernamonti:2020bcf,Bai:2021ldj}.

Motivated by these considerations, it is clearly interesting to consider the computation of complexity in conformal field 
theory and a number of interesting results have already been achieved.
For free and weakly coupled quantum field theories, using the language of quantum circuits, complexity becomes the length of the shortest geodesic in the space of circuits \cite{Jefferson:2017sdb,Chapman:2017rqy,Khan:2018rzm,Hackl:2018ptj,Chapman:2018hou,Chapman:2019clq,Bhattacharyya:2018bbv,Jiang:2018nzg,Doroudiani:2019llj,Caceres:2019pgf,Guo:2020dsi,Meng:2021wmz,Moghimnejad:2021rqe,Chapman:2017rqy,Yang:2017czx,Sinamuli:2019utz}.
This leads to a deep connection between complexity and geometry\cite{Brown:2019whu,Auzzi:2020idm}.
Another setting in which complexity has been studied, is for two dimensional conformal field theories, which enjoy an 
enhanced (infinite dimensional) symmetry group that can effectively be used \cite{Caputa:2018kdj,Erdmenger:2020sup,Flory:2020eot,Flory:2020dja,Bueno:2019ajd}.
The two dimensional case is however special and it is clearly interesting to generalize the evaluation of computational 
complexity  to higher dimensional conformal field theories.
The paper \cite{Chagnet:2021uvi} has outlined a practical approach to this problem.
The idea is to use the conformal symmetry generators to define simple unitary operators (simple gates).
These gates are then the building blocks of quantum circuits, defined in a unitary representation of the conformal group.
The resulting unitary evolution defines a protocol whose computational cost is given by a suitable notion of length in 
the unitary manifold of conformal symmetries.
In general computational costs can be used to penalize or reward certain gates.
The choice of cost function is not unique and the answer for the computed complexity depends on the choice of 
the computational cost.
Assuming that all symmetry transformations are equally easy to perform, the cost function for the class of simple gates we 
consider is fixed up to a global choice of units\cite{Magan:2018nmu}. 
Using these circuits, \cite{Chagnet:2021uvi} present explicit results for state dependent distance functions along such 
circuits, starting from a scalar primary state.
The circuits live in a phase space which allows an intimate connection to coadjoint orbits from representation theory. 
The coadjoint orbit can be identified with the coset space SO($d$, 2)/SO(2)$\times$SO($d$), and the geometry of the
coadjoint orbit reproduces the results of the quantum circuit analysis.

In this paper we generalize the results of \cite{Chagnet:2021uvi} to circuits which start from arbitrary spinning primary states.
We work in $d=3$ and $d=4$ dimensions, where the relevant conformal groups are SO(2,3) and SO(2,4).
One of our main results are explicit state dependent distance functions along circuits which start from primary states with
an arbitrary spin.
These results are again in complete agreement with results obtained by analyzing the geometry of the relevant coadjoint
orbits.
There are a number of features of our final formulas that are noteworthy.
In the quantum circuit approach, using the state operator correspondence of radial quantization, the distance functions 
which compute complexity are given in terms of matrix elements which can be related to specific two point functions of
spinning primaries.
This establishes a precise link between complexity and correlation functions of the conformal field theory.
Since the form of two point functions are completely determined by conformal symmetry, we obtain concrete formulas
for the relevant matrix elements.
Working in the basis which diagonalizes suitably chosen rotation generators in the chosen Cartan sub algebra, we 
exhibit a remarkably simple dependence on the spin of the primary.
A completely independent approach to these matrix elements, using the Baker-Campbell-Hausdorff formulas, confirms
these conclusions.

The paper is organized as follows: in Section \ref{threedim} we study circuits in unitary representations of the three
dimensional Lorentzian conformal group SO(2,3).
In the language of radial quantization the bulk of the computation entails evaluating matrix elements of the unitary operator
defining the circuit.
The computation is described in some detail because the structure of the result has direct application to four dimensional
conformal field theories, with conformal group SO(2,4).
This is described in Section \ref{fourdim}.
We confirm our conformal field theory results using a geometrical analysis of the relevant coadjoint orbits.
In Section \ref{conclusions} we discuss our results and describe some avenues that can be pursued.
The Appendix \ref{BCH} collects results that are useful for the application of the Baker-Campbell-Hausdorff formula,
while Appendix \ref{MG} discusses the dependence of the complexity geometry on the reference state chosen.

\section{Complexity from $SO(2,3)$ for a spining primary reference state}\label{threedim}

The goal of this section is to compute complexity by considering a quantum circuit that starts from a spinning 
primary reference state.
The construction of the circuit employs Euclidean conformal generators, which obey an SO(1,$d$+1) algebra. 
Following \cite{Chagnet:2021uvi} our choice of Hermiticity conditions ensures that we are building unitary circuits of 
the Lorentzian conformal group SO(2,$d$). 
For more details the reader should consult \cite{Minwalla:1997ka,Luscher:1974ez,Chagnet:2021uvi}. 

\subsection{Conformal Algebra}

Following the conventions of \cite{Chagnet:2021uvi}, the Euclidean algebra is ($\mu,\nu,\rho,\sigma=1,2,3$)
\bea
[D,P_\mu]&=&P_\mu\qquad
[D,K_\mu]=-K_\mu\cr
[K_\mu,P_\nu]&=&2(\delta_{\mu\nu}D-L_{\mu\nu})\cr
[L_{\mu\nu},P_\rho]&=&\delta_{\nu\rho}P_\mu-\delta_{\mu\rho}P_\nu\qquad
[L_{\mu\nu},K_\rho]=\delta_{\nu\rho}K_\mu-\delta_{\mu\rho}K_\nu\cr
[L_{\mu\nu},L_{\rho\sigma}]&=&\delta_{\nu\rho}L_{\mu\sigma}-\delta_{\mu\rho}L_{\nu\sigma}
-\delta_{\nu\sigma}L_{\mu\rho}+\delta_{\mu\sigma}L_{\nu\rho}
\eea
together with the hermitticity conditions
\bea
D^\dagger=D\qquad K_\mu^\dagger=P_\mu\qquad L_{\mu\nu}^\dagger=-L_{\mu\nu}
\eea
The $L_{\mu\nu}$'s are anti-hermitian generators of SO$(3)$.
After removing a factor of $i$ we find hermitian generators
\bea
J_a={1\over 2i}\epsilon_{abc}L_{bc}
\eea
which obey the usual algebra
\bea
[J_a,J_b]=i\epsilon_{abc}J_c
\eea
Raising and lowering operators are given by the standard formulas
\bea
J_\pm=J_1\pm iJ_2
\eea
The highest weight state obeys
\bea
J_+|jj\rangle =0\qquad J_3|jj\rangle =j|jj\rangle
\eea
Finally a well known but useful formula for the analysis below is
\bea
J_-|jm\rangle =\sqrt{(j+m)(j-m+1)}|j\, m-1\rangle
\eea

\subsection{Reference State}

Our circuit will start from a reference state $|\Delta,j;j\rangle$, which is a primary of dimension $\Delta$ and a highest 
weight state of the spin $j$ multiplet.
Consequently we have
\bea
D|\Delta,j;j\rangle &=&\Delta |\Delta,j;j\rangle \qquad 
K_\mu|\Delta,j;j\rangle =0\cr\cr
 J_3|\Delta,j;j\rangle&=& j|\Delta,j;j\rangle\qquad J_+ |\Delta,j;j\rangle=0
\eea

\subsection{Allowed Gates}

Circuits are generated using the unitary
\begin{equation}
U(\sigma)=e^{i\alpha\cdot P}e^{i\gamma D}\left(e^{i\lambda_{-}J_-}
e^{i\lambda_{3}J_3}e^{i\lambda_{+}J_+}\right)
e^{i\beta\cdot K}
\end{equation}
The coordinates $\alpha^\mu$, $\gamma$, $\lambda_{\pm}$, $\lambda_3$ and $\beta^\mu$ are all functions of $\sigma$.
When acting on our reference state with this primary that are a number of immediate simplifications: our target state is primary
so it is annihilated by $K_\mu$ and it is highest weight state, so it is annihilated by $J_+$.
Consequently we have
\begin{eqnarray}
\left(e^{i\lambda_{-}J_-} e^{i\lambda_{3}J_3}e^{i\lambda_{+}J_+}\right)
e^{i\beta\cdot K}|\Delta,j;j\rangle
&=&e^{i\lambda_{3}j}e^{i\lambda_{-}J_-}|\Delta,j;j\rangle\cr\cr
&=&e^{i\lambda_{3}j}
\sum_{n=0}^{2j}(i\lambda_-)^n c_{n} |\Delta,j;j-n\rangle
\end{eqnarray}
where $c_0=1$ and
\bea
c_{n}=\prod_{k=0}^{n-1}\sqrt{2j-k\over k+1}
=\sqrt{2j\choose n}\qquad n\ge 1
\eea
Thus, allowing $U(\sigma)$ to act on our target state, we find
\bea
U(\sigma)|\Delta,j;j\rangle =
e^{i\gamma \Delta+i\lambda_{3}j}\sum_{n=0}^{2j} (i\lambda_-)^{n} c_{n}e^{i\alpha\cdot P}|\Delta,j;j-n\rangle
\equiv |\alpha\rangle\label{alphastate}
\eea

\subsection{Cost Function and Fubini-Study Metric}\label{costone}

To define computational complexity, we have a (typically simple) reference state and a target state.
We have a task which entails acting on the reference state with a product of unitary operators (chosen from a set of simple 
gates) to produce the target state.
Complexity is then defined as the minimal number of gates required to achieve the task.
The discrete phrasing of complexity is not very useful in a field theory setting where we are dealing with systems that have
an infinite number of degrees of freedom.
A continuous notion of complexity is more natural.
To achieve this, we can replace gate counting with the problem of computing geodesic lengths on the manifold of unitary
operators used to define the quantum circuit.
The circuit inherits a continuous parameter $\sigma$.
As $\sigma$ is advanced from zero, the circuit takes the reference state into the target state.
An evolution from $\sigma$ to $\sigma+d\sigma$ corresponds to an infinitesimally short path.
We use a cost function to assign a length to this short path, and then compute lengths of finite paths by integration.

Following \cite{Caputa:2018kdj} we consider two distance functions: an ${\cal F}_1$ cost function and the Fubini-Study norm.
The cost function, which computes the length of an infinitesimally short path, is computing the norm of a tangent to the path. 
Tangents to paths on the manifold of unitaries are valued in the Lie algebra, so it's natural to consider a cost
$\propto  U^\dagger dU$.
The ${\cal F}_1$ cost function gives a number by considering the expected value of $U^\dagger dU$ as follows
\begin{equation}
{\cal F}_1 =|\langle\Delta,j;j|U^\dagger dU|\Delta,j;j\rangle|
\end{equation}
As argued in \cite{Chagnet:2021uvi}, the ${\cal F}_1$ cost function assigns zero cost to certain gates, which limits its
effectiveness as a complexity measure. 
For this reason we focus on the closely related measure, the Fubini-Study norm
\begin{equation}
ds^2=\langle\Delta,j;j |dU^\dagger dU|\Delta,j;j\rangle -|\langle\Delta,j;j|U^\dagger dU|\Delta,j;j\rangle|^2
\end{equation}
We will argue, by explicit computation, that the resulting Fubini-Study metric is a positive definite Einstein-K\"ahler metric.
Further, our results can be understood in terms of the geometry of coadjoint orbits, in line with the discussion 
of \cite{Caputa:2018kdj}.
A straight forward computation shows that
\bea
ds^2&=&\Big( \langle\alpha|\alpha^*\cdot K\alpha\cdot P|\alpha\rangle
-\langle\alpha|\alpha^*\cdot K|\alpha\rangle\langle\alpha|\alpha\cdot P|\alpha\rangle
+\langle\beta|\beta\rangle-\langle\beta|\alpha\rangle\langle\alpha|\beta\rangle\cr\cr
&&-i\langle\alpha|\alpha^*\cdot K|\beta\rangle+i\langle\alpha|\alpha^*\cdot K|\alpha\rangle\langle\alpha|\beta\rangle
+i\langle\beta|\alpha\cdot P|\alpha\rangle-i\langle\alpha|\alpha\cdot P|\alpha\rangle\langle\beta|\alpha\rangle\Big) d\sigma^2
\cr
&&
\eea
where
\bea
|\alpha\rangle=\sum_{m=0}^{2j}e^{i\gamma\Delta+i\lambda_3 j}(i\lambda_-)^m c_m
e^{i\alpha\cdot P}|\Delta,j;j\rangle
\eea
\bea
|\beta\rangle=\sum_{m=0}^{2j}im\dot{\lambda}_- e^{i\gamma\Delta+i\lambda_3 j}(i\lambda_-)^{m-1} c_m
e^{i\alpha\cdot P}|\Delta,j;j\rangle
\eea
To evaluate this metric, we need to evaluate some expectation values.
This is performed in detail in the next section.
Freely using the relevant results, it is now straight forward to obtain the Fubini-Study metric
\bea
ds^2&=&2 (\Delta +j) \left(
\frac{2 (\alpha^*\cdot d\alpha-\alpha^*\cdot\alpha^*\alpha \cdot d\alpha) 
(\alpha\cdot d\alpha^* -\alpha\cdot\alpha  \alpha^*\cdot d\alpha^*)}{(1-2 \alpha .\alpha^*+|\alpha\cdot\alpha|^2)^2}
+\frac{d\alpha^*\cdot d\alpha-2 \alpha\cdot d\alpha  \alpha^*\cdot d\alpha^*}
{1-2 \alpha\cdot\alpha^*+|\alpha\cdot\alpha|^2}\right)\cr\cr\cr
&-&2 j \left(\frac{d\alpha^*\cdot M\cdot \alpha  \alpha^*\cdot M\cdot d\alpha}{\Lambda^2}
+\frac{d\alpha^*\cdot M\cdot d\alpha}{\Lambda}\right)
+2j\,\, \frac{1-2 \alpha \cdot\alpha^*+|\alpha\cdot\alpha|^2}{\Lambda^2}\,\,d\lambda_- d\lambda_-^*\cr\cr\cr
&+&2j\left[
\frac{i d\lambda_-(\alpha\cdot L^*\cdot d\alpha^*-\alpha\cdot\alpha  \alpha^*\cdot L^*\cdot d\alpha^*)}
{\Lambda^2}
+\frac{i d\lambda_-^* (\alpha^*\cdot L\cdot d\alpha-\alpha^*\cdot\alpha^* \alpha\cdot L\cdot d\alpha)}{\Lambda^2}
\right]\label{FSMetric}
\eea
where
\bea
M=\left[
\begin{array}{ccc}
 -\lambda_- \lambda_-^*-1 & -i+i \lambda_- \lambda_-^* & i (\lambda_-+\lambda_-^*) \\
 -i (\lambda_- \lambda_-^*-1) & -\lambda_- \lambda_-^*-1 & \lambda_--\lambda_-^* \\
 -i (\lambda_-+\lambda_-^*) & \lambda_-^*-\lambda_- & -\lambda_- \lambda_-^*-1 \\
\end{array}
\right]\label{deffM}
\eea
\bea
L=\left(
\begin{array}{ccc}
 0 & 2 \lambda_- & 1-\lambda_-^2 \\
 -2 \lambda_- & 0 & i \left(\lambda_-^2+1\right) \\
 \lambda_-^2-1 & -i \left(\lambda_-^2+1\right) & 0 \\
\end{array}
\right)\label{deffL}
\eea
and
\bea
\Lambda&=&-((1-\alpha\cdot \alpha^*) (\lambda_- \lambda_-^*+1)-i (\lambda_- \lambda_-^*-1) 
(\alpha_1 \alpha_2^*-\alpha_2 \alpha_1^*)\cr
&&-i (\lambda_-+\lambda_-^*) (\alpha_1 \alpha_3^*-\alpha_3 \alpha_1^*)
+(\lambda_--\lambda_-^*) (\alpha_3 \alpha_2^*-\alpha_2 \alpha_3^*))\label{forp}
\eea
The result (\ref{FSMetric}) is one of the key new results of this paper.

The stability group of a scalar primary is SO(2)$\times$SO(3).
In this case the metric is defined on a 6 dimensional space with complex coordinates $\alpha^\mu$,
which can be identified with the coset space SO(2,3)/SO(2)$\times$SO(3) \cite{Chagnet:2021uvi,Gibbons:1999rb}.
The stability group of a spinning primary is SO(2)$\times$SO(2).
In this case the metric is defined on an 8 dimensional space with complex coordinates $\alpha^\mu$, $\lambda_-$.
This space is the coset space SO(2,3)/SO(2)$\times$SO(2).

\subsection{Some Expectation Values}

In this section we will compute the matrix elements need to evaluate the Fubini-Study metric.
Using the state operator correspondence of radial quantization, the matrix elements we need can be related to specific 
two point functions of primary operators.
This is a useful observation because primary two point functions are completely determined by conformal symmetry.
Working in the basis which diagonalizes $J_3$ and $\vec{J}\cdot\vec{J}$, we find a remarkably simple dependence on 
the spin of the primary.
These results are in complete agreement with an independent evaluation making use of the Baker-Campbell-Hausdorff formulas.
It is useful to start with a study of spin zero primaries.
The operator state correspondence implies the following identity
\bea
\langle \Delta |e^{-i\alpha^*\cdot K}e^{i\alpha\cdot P}|\Delta\rangle
=\langle I{\cal O}_\Delta(y')I{\cal O}_\Delta(x)\rangle
\eea
between matrix elements and two point functions.
The two point function is
\bea
\langle {\cal O}_\Delta (y){\cal O}_\Delta(x)\rangle &=& {1\over |x-y|^{2\Delta}}={1\over (x^2-2x\cdot y+y^2)^\Delta}\cr\cr
&=&{1\over (y^2)^\Delta}{1\over (x^2 y^{\prime 2}-2x\cdot y'+1)^\Delta}
\eea
which implies that
\bea
\langle I{\cal O}_\Delta(y')I{\cal O}_\Delta(x)\rangle = {1\over (x^2 y^{\prime 2}-2x\cdot y'+1)^\Delta}
\eea
Thus, the matrix element we want is 
\bea
\langle \Delta |e^{-i\alpha^*\cdot K}e^{i\alpha\cdot P}|\Delta\rangle
={1\over (\alpha^2\alpha^{* 2}-2\alpha\cdot\alpha^*+1)^{\Delta}}
\eea

The above observation generalizes to spinning primary operators.
The simplest spinning example we can consider is a spin one primary of dimension $\Delta$, denoted 
${\cal O}_{\Delta,\mu}(x)$.
In this case the state operator correspondence implies that
\bea
\langle \Delta,\mu|e^{-i\alpha^*\cdot K}e^{i\alpha\cdot P}|\Delta,\nu\rangle
=\langle I{\cal O}_{\Delta,\mu}(y')I{\cal O}_{\Delta,\nu}(x)\rangle
\eea
Using \cite{Osborn:1993cr}
\bea
I{\cal O}_{\Delta,\nu}(x')I=x^{2\Delta}\left(\delta_{\mu\nu}-2{x'_\mu x'_\nu\over x^{\prime 2}}\right)O_{\Delta,\mu}(x)
\eea
as well as the two point function of the spinning primary \cite{Osborn:1993cr}
\bea
\langle{\cal O}_{\Delta,\mu}(x){\cal O}_{\Delta,\nu}(y)\rangle
={\delta_{\mu\nu}-2{(x-y)_\mu(x-y)_\nu\over (x-y)^2}\over |x-y|^{2\Delta}}
\eea
we find
\bea
\langle I{\cal O}_{\Delta,\mu}(y')I{\cal O}_{\Delta,\nu}(x)\rangle 
&=&\left(\delta_{\mu}^{\alpha}-2{(y')_\mu(y')^\alpha\over (y')^2}\right)
{\delta_{\alpha\nu}-2{(x-y)_\alpha(x-y)_\nu\over (x-y)^2}\over (x^2 y^{\prime 2}-2x\cdot y'+1)^\Delta}\cr\cr\cr
&=&\left(\delta_{\mu}^{\alpha}-2{(y')_\mu(y')^\alpha\over (y')^2}\right)
{\delta_{\alpha\nu}-2{(y^{\prime 2}x-y')_\alpha(y^{\prime 2}x-y')_\nu\over y^{\prime 2}(1-2x\cdot y'+y^{\prime 2}x^2)}
\over (x^2 y^{\prime 2}-2x\cdot y'+1)^\Delta}
\eea
This spin 1 example generalizes easily to higher spins.
As a final example, we consider the spin 2 primary $O_{\Delta,\mu\nu}(x)$.
This operator is symmetric and traceless on the $\mu\nu$ indices.
The matrix element we wish to compute is
\bea
\langle \Delta,\alpha\beta|e^{-i\alpha^*\cdot K}e^{i\alpha\cdot P}|\Delta,\mu\nu\rangle
=\langle I{\cal O}_{\Delta,\alpha\beta}(y')I{\cal O}_{\Delta,\sigma\tau}(x)\rangle
\eea
Using the action of inversion \cite{Osborn:1993cr}
\bea
IO_{\Delta,\mu\nu}(x)I&=&{(x')^{2\Delta}\over 2}
\left(I_{\mu\alpha}(x')I_{\nu\beta}(x')+I_{\nu\alpha}(x')I_{\mu\beta}(x')\right)
{\cal O}_{\Delta,\alpha\beta}(x')\cr\cr
&=&(x')^{2\Delta} I_{\mu\alpha}(x')I_{\nu\beta}(x') {\cal O}_{\Delta,\alpha\beta}(x')
\eea
where
\bea
I_{\mu\nu}(x')=\left(\delta_{\mu\nu}-2{x'_\mu x'_\nu\over x^{\prime 2}}\right)
\eea
and the two point function \cite{Osborn:1993cr}
\bea
\langle{\cal O}_{\Delta,\mu\nu}(x){\cal O}_{\Delta,\alpha\beta}(y)\rangle
={I_{\mu\alpha}(x-y)I_{\nu\beta}(x-y)+I_{\nu\alpha}(x-y)I_{\mu\beta}(x-y)
-{2\over d}\delta_{\mu\nu}\delta_{\alpha\beta}
\over 2|x-y|^{2\Delta}}
\eea
we find
\bea
&&\langle I{\cal O}_{\Delta,\alpha\beta}(y')I{\cal O}_{\Delta,\sigma\tau}(x)\rangle 
=I_{\alpha\mu}(x')I_{\beta\nu}(x')\cr\cr
&&\qquad\qquad\times
{I_{\mu\sigma}(x-y)I_{\nu\tau}(x-y)+I_{\nu\sigma}(x-y)I_{\mu\tau}(x-y)
-{2\over d}\delta_{\mu\nu}\delta_{\sigma\tau}
\over 2(x^2 y^{\prime 2}-2x\cdot y'+1)^\Delta}
\eea
The generalization to arbitrary spin is now obvious.

The above argument demonstrates that the matrix elements relevant for the computation of complexity can be expressed in
terms of correlation functions.
Since the holographic computation of correlation functions is well developed, this connection may well find application in
the computation of complexity at strong coupling in the conformal field theory.

The results obtained above can be simplified dramatically to make the dependence on the spin of the primary 
operator transparent.
The discussion above expressed primary operators in terms of traceless symmetric tensors.
There is an alternative description which uses states with definite $J_3$ and $\vec{J}\cdot\vec{J}$ quantum numbers.
This description makes it simple to evaluate the action of the so(3) raising and lowering operators on the states and will
ultimately exhibit the very simple dependence on the spin $j$ of the primary. 
The change of basis from the $|\Delta,\mu_1\cdots\mu_j\rangle$ to the $|\Delta,j;m\rangle$ is the transformation from 
traceless symmetric tensors to spherical harmonics.
Concretely, any function of the angles $\theta,\phi$ on the unit sphere can be expanded in two ways
\bea
F(\theta,\phi)&=&\sum_{l=0}^\infty c^{(l)}_{\mu_1\mu_2\cdots\mu_l}x^{\mu_1}x^{\mu_2}\cdots x^{\mu_l}\cr\cr
&=&\sum_{l=0}^\infty\sum_{m=-l}^l a_{lm}Y_{lm}(\theta,\phi)
\eea
where $x^\mu$ is a coordinate on the unit sphere, so that
\bea
x^1=\sin\theta\cos\phi\qquad
x^2=\sin\theta\sin\phi\qquad
x^3=\cos\theta
\eea
The relation between the two bases takes the form
\bea
C^{(l,m)}_{\mu_1\cdots\mu_l}x^{\mu_1}\cdots x^{\mu_l} =Y_{lm}(\theta,\phi)\label{SphH}
\eea
We can use the coefficients of this change of basis to transform states
\bea
|\Delta,j;m\rangle =C^{(j,m)}_{\mu_1\cdots\mu_j}|\Delta,j;\mu_1\cdots\mu_l\rangle
\eea
or to transform operators
\bea
{\cal O}_{\Delta,j;m} =C^{(j,m)}_{\mu_1\cdots\mu_j}{\cal O}_{\Delta,j;\mu_1\cdots\mu_l}
\eea
Using (\ref{SphH}) we can read off the coefficients for the change of basis $C^{(j,m)}_{\mu_1\cdots\mu_j}$ directly
from the spherical harmonics.
The only tricky thing about this procedure is that traces must be subtracted off by hand
\bea
C^{(j,m)}_{\mu_1\cdots \mu_j}={\partial\over\partial x^{\mu_1}}\cdots{\partial\over\partial x^{\mu_j}}
Y_{jm}(\theta,\phi)-{\rm traces}
\eea
This can efficiently be dealt with by employing the Thomas derivative
\bea
D_\mu=\left(h-1+x\cdot {\partial\over\partial x}\right){\partial\over\partial x^\mu}-{1\over 2}x_\mu 
{\partial\over\partial x}\cdot {\partial\over\partial x}
\eea
with $h={d\over 2}={3\over 2}$.
Acting on any polynomial $D_\mu$ automatically subtracts the traces off, so that
\bea
C^{(j,m)}_{\mu_1\cdots \mu_j}=D_{\mu_1}\cdots D_{\mu_j}Y_{jm}(\theta,\phi)
\eea
This gives the change of basis for any spin $j$.
We will use the primary that is the highest weight state of the spin $j$ multiplet.
The highest weight state has $m=j$ so that a formula that will prove to be useful in what follows is
\bea
Y_{jj}(\theta,\phi)\propto (x^1+ix^2)^j\label{sphharm1}
\eea
where we will not need the overall normalization.
Applying the Thomas derivative, we easily find
\bea
C^{(j,j)}_{\mu_1\cdots \mu_j}\propto (\delta_{\mu_1 1}+i\delta_{\mu_1 2})
(\delta_{\mu_2 1}+i\delta_{\mu_2 2})\cdots (\delta_{\mu_j 1}+i\delta_{\mu_j 2})\label{bchange}
\eea
This product structure is a consequence of the fact that the highest weight state $|j,j\rangle$ can be written as the tensor
product of $j$ copies of the state $|1,1\rangle$
\bea
|j,j\rangle &=&|1,1\rangle\otimes |1,1\rangle\otimes\cdots\otimes |1,1\rangle\cr\cr
&=&|1,1\rangle^{\otimes\, j}
\eea
The action of the Lie algebra on a tensor product is through the usual co-product
\bea
J_- |j,j\rangle &=&(J_-|1,1\rangle)\otimes |1,1\rangle\otimes\cdots\otimes |1,1\rangle
+|1,1\rangle\otimes (J_- |1,1\rangle)\otimes\cdots\otimes |1,1\rangle\cr\cr
&+&\cdots+|1,1\rangle\otimes |1,1\rangle\otimes\cdots\otimes (J_-|1,1\rangle)
\eea
It now easily follows that
\bea
e^{i\lambda_- J_-}|j,j\rangle &=&(e^{i\lambda_- J_-}|1,1\rangle)\otimes (e^{i\lambda_- J_-}|1,1\rangle)
\otimes\cdots\otimes (e^{i\lambda_- J_-}|1,1\rangle)\label{prodstruct}
\eea
Thus, the product structure enjoyed by the highest weight state continues for $e^{i\lambda_- J_-}|j,j\rangle$.
This has immediate application when computing matrix elements using the state $|\alpha\rangle$ defined in
(\ref{alphastate}).
A key result is
\bea
&&\langle\Delta,j;j|e^{-i\lambda_-^*J_+}e^{-i\alpha^*\cdot K}
e^{i\alpha\cdot P}e^{i\lambda_-J_-}|\Delta,j;j\rangle\cr\cr
&=&C^{(j,j)}_{\mu_1\cdots\mu_j}C^{(j,j)}_{\nu_1\cdots\nu_j}
\langle\Delta,\nu_1\cdots\nu_j|e^{-i\lambda_-^*J_+}e^{-i\alpha^*\cdot K}
e^{i\alpha\cdot P}e^{i\lambda_-J_-}|\Delta,\mu_1\cdots\mu_2\rangle\cr\cr
&=&{\Lambda^{2j}\over
(1-2\alpha\cdot\alpha^*+\alpha\cdot\alpha\alpha^*\cdot\alpha^*)^{\Delta+j}}\label{GenRes}
\eea
where $\Lambda$ was defined in \eqref{forp}.
The form given in the last line of (\ref{GenRes}) follows using (\ref{prodstruct}), and the specific form of the
polynomial given in  (\ref{forp}) can be read from the $j=1$ result.
Note the remarkably simple dependence of the last line of \eqref{GenRes} on the spin $j$ of the primary.
The final result (\ref{GenRes}) can be recovered in a completely independent way, by using the Baker-Campbell-Hausdorff
formula to evaluate the first line directly.
See Appendix \ref{BCH} for the details.

Using the above result we easily find
\bea
\langle\alpha|\alpha\rangle
=e^{i(\gamma-\gamma^*)\Delta+i(\lambda_3-\lambda_3^*)j}
{\Lambda^{2j}\over
(1-2\alpha\cdot\alpha^*+\alpha\cdot\alpha\alpha^*\cdot\alpha^*)^{\Delta+j}}
\label{SO23Result}
\eea
Derivatives of this last expression, with respect to the appropriate parameters, gives the required matrix elements.
For example
\bea
\langle\alpha|P_\mu |\alpha\rangle
=-i{\partial\over\partial \alpha_\mu}\log\left(
{\Lambda^{2j}\over
(1-2\alpha\cdot\alpha^*+\alpha\cdot\alpha\alpha^*\cdot\alpha^*)^{\Delta+j}}
\right)
\eea
and
\bea
\langle\alpha|K_\mu |\alpha\rangle
=i{\partial\over\partial \alpha^*_\mu}\log\left(
{\Lambda^{2j}\over
(1-2\alpha\cdot\alpha^*+\alpha\cdot\alpha\alpha^*\cdot\alpha^*)^{\Delta+j}}
\right)
\eea
Setting $j=0$ and evaluating the above two expression, we find complete agreement with the results given 
in \cite{Chagnet:2021uvi}.

\subsection{Fubini-Study metric from coadjoint orbit}\label{coadjoint}

The above result for the Fubini-Study metric can be reproduced by studying coadjoint orbits of the conformal group.
Denote the defining representation by $R(\cdot)$.
The generators of so(2,$d$) in the defining representation are given by
\begin{equation}
(M_{AB})^C{}_D=\delta_A{}^Cg_{BD}-\delta_B{}^Cg_{AD}\qquad A,B,C,D=-1,0,1,...,d
\end{equation}
where the $d+2$ dimensional metric $g={\rm diag}(-,-,+,+,\cdots,+)$.
The elements of the conformal algebra are represented as follows
\begin{eqnarray}
&&R(D)=-iM_{-1,0}\qquad\qquad\quad R(L_{\mu\nu})=M_{\mu\nu}\cr\cr
&&R(P_\mu)=M_{-1,\mu}-iM_{0,\mu}\qquad
R(K_\mu)=-(M_{-1,\mu}+iM_{0,\mu})
\end{eqnarray}
We can now construct $R(U)$, with
\bea
U(\sigma )=e^{-i\alpha\cdot P}e^{i\gamma D}e^{i\lambda^{\mu\nu}L_{\mu\nu}}e^{i\beta\cdot K}
\eea
The parameters $\alpha^\mu,\gamma,\lambda^{\mu\nu},\beta$ are all complex numbers.
Using the so(2,3) algebra and requiring that $R(U)$ is unitary, the above expression simplifies to
\bea
U &=& e^{i \alpha \cdot P} e^{i \sigma \cdot K} e^{ i\gamma_R D +  \log\left( \gamma \right) D} e^{i \lambda_{-} J_{-}} e^{i \lambda_{+} J_{+}} e^{ i \lambda_{3 R} J_3 + \left( \log(\gamma) -\log(\Lambda) \right) J_3}
\eea
with parameters
\bea
\sigma_{u} = -\frac{\partial}{\partial \alpha^{\mu}} \log\left( \gamma \right)\qquad
\lambda_{+}=\frac{\partial}{\partial \lambda_{-}} \log\left( \Lambda \right) \qquad
\gamma =\sqrt{1 - 2 \alpha \cdot \alpha^* + |\alpha|^4} \label{defgam}
\eea
and where $\Lambda$ is defined in \eqref{forp}.
Thus, after imposing unitarity, the variables parametrizing the unitary $U(\sigma)$ are $\alpha^\mu$ which is a complex 
vector with three components,
$\gamma_R$ a real number, $\lambda_{3R}$ a real number and $\lambda_-$ which is a complex number.
This is a total of 10 real numbers which matches the dimension of so(2,3).
It is now possible to compute the Maurer-Cartan form
\bea
\Theta =U^{-1}dU
\eea
In terms of the Maurer-Cartan form, we can compute the symplectic form
\bea
{\rm Tr}(\eta \cdot dR(\Theta))=\omega_{\bar{\mu}\nu} d\alpha^{*\bar{\mu}}\wedge d\alpha^\nu
\qquad R(\Theta)=R(U)^{-1}dR(U)
\eea
where we have used the following element
\bea
\eta ={\Delta\over 2} M_{-1,0}- {j\over 2} M_{1,2}
\eea
The stabilizer algebra of $\eta$ is so(2)$\times$so(2).
The choice of $\eta$ is made to match the stabilizer algebra of the spinning primary reference state.
The metric on the orbit is then obtained by contracting with the complex structure
\begin{equation}
J^\mu{}_\rho d\alpha^\rho=-id\alpha^\mu \qquad 
J^{\bar\mu}{}_{\bar\rho} d\alpha^{\bar\rho}=id\alpha^{\bar\mu}
\end{equation}
to yield
\begin{equation}
ds^2=-i\omega_{\bar{\mu}\nu}d\alpha^{*\bar\mu}d\alpha^\nu
\end{equation}
The result is in complete agreement with \eqref{FSMetric}.

\section{Complexity from $SO(2,4)$ for a spining primary reference state}\label{fourdim}

The quantum circuits considered in this section are in unitary representations of the Lorentzian conformal group SO(2,4).
We will see that the computation is a simple generalization of the computation given in Section \ref{threedim}.
In particular, the matrix elements needed to evaluate the Fubini-Study metric can again be related to two point correlation
functions and final results again have a remarkably simple dependence on the spin quantum numbers of the primary
reference state.
Once again, the final results also follow from an independent evaluation which employs the Baker-Campbell-Hausdorff formula. 

\subsection{Conformal Algebra}

Again following the conventions of \cite{Chagnet:2021uvi}, the Euclidean algebra is ($\mu,\nu,\rho,\sigma=1,2,3,4$)
\bea
[D,P_\mu]&=&P_\mu\qquad
[D,K_\mu]=-K_\mu\cr
[K_\mu,P_\nu]&=&2(\delta_{\mu\nu}D-L_{\mu\nu})\cr
[L_{\mu\nu},P_\rho]&=&\delta_{\nu\rho}P_\mu-\delta_{\mu\rho}P_\nu\qquad
[L_{\mu\nu},K_\rho]=\delta_{\nu\rho}K_\mu-\delta_{\mu\rho}K_\nu\cr
[L_{\mu\nu},L_{\rho\sigma}]&=&\delta_{\nu\rho}L_{\mu\sigma}-\delta_{\mu\rho}L_{\nu\sigma}
-\delta_{\nu\sigma}L_{\mu\rho}+\delta_{\mu\sigma}L_{\nu\rho}
\eea
This algebra is SO(1,5).
We again impose Hermiticity conditions
\bea
D^\dagger=D\qquad K_\mu^\dagger=P_\mu\qquad L_{\mu\nu}^\dagger=-L_{\mu\nu}
\eea
to ensure that our circuit gives a unitary representations of the Lorentzian conformal group SO(2,4).
Here the $L_{\mu\nu}$'s are antihermittian generators of SO$(4)$.
As usual, we rewrite SO(4) as SU(2)$\times$SU(2) as follows (these are Hermittian generators)
\bea
J_i^R={1\over 2i}\left(L_{4i}+{1\over 2}\epsilon_{ijk}L_{jk}\right)\qquad\qquad
J_i^L={i\over 2}\left(L_{4i}-{1\over 2}\epsilon_{ijk}L_{jk}\right)
\eea
These obey the algebra (there is a sum on $k$ on the RHS and $\epsilon_{123}=1$)
\bea
[J_i^R,J_j^R]=i\epsilon_{ijk}J^R_k\qquad
[J_i^L,J_j^L]=i\epsilon_{ijk}J^L_k\qquad
[J_i^R,J_j^L]=0
\eea
In a basis of $J_3^R$ and $J^L_3$ we have
\bea
J_3^R |j_R,j_L;m_R,m_L\rangle =m_R |j_R,j_L;m_R,m_L\rangle\qquad
J_3^L |j_R,j_L;m_R,m_L\rangle =m_L |j_R,j_L;m_R,m_L\rangle\nonumber
\eea
The raising and lowering operators are given as usual by
\bea
J_\pm^R=J_1^R\pm iJ_2^R\qquad J_\pm^L=J_1^L\pm iJ_2^L
\eea
We will use a reference state that is a highest weight state for both ($R$ and $L$) SU(2)s.

\subsection{Reference State}

The reference state $|\Delta,j_R,j_L;j_R,j_L\rangle$ is a primary of dimension $\Delta$ and a highest weight state of the
$(j_R,j_L)$ multiplet, so that
\bea
D|\Delta,j_R,j_L;j_R,j_L\rangle &=&\Delta |\Delta,j_R,j_L;j_R,j_L\rangle \qquad 
K_\mu|\Delta,j_R,j_L;j_R,j_L\rangle =0\cr\cr
 J_3^R|\Delta,j_R,j_L;j_R,j_L\rangle&=& j_R|\Delta,j_R,j_L;j_R,j_L\rangle\qquad
J_3^L|\Delta,j_R,j_L;j_R,j_L\rangle =j_L|\Delta,j_R,j_L;j_R,j_L\rangle\cr\cr
J_+^R |\Delta,j_R,j_L;j_R,j_L\rangle&=&0\qquad J_+^L |\Delta,j_R,j_L;j_R,j_L;j_R,j_L\rangle=0
\eea

\subsection{Allowed Gates}

Circuits are generated using the unitary
\begin{equation}
U(\sigma)=e^{i\alpha\cdot P}e^{i\gamma D}\left(e^{i\lambda^R_{-}J_-^R +i\lambda^L_{-}J_-^L}
e^{i\lambda^R_{3}J_3^R +i\lambda^L_{3}J_3^L}e^{i\lambda^R_{+}J_+^R +i\lambda^L_{+}J_+^L}\right)
e^{i\beta\cdot K}
\end{equation}
The coordinates $\alpha^\mu$, $\gamma_D$, $\lambda_{*}^R$, $\lambda_{*}^L$ and $\beta^\mu$ are all 
functions of $\sigma$.
The action of this unitary on our reference state simplifies nicely.
Using the fact the reference state is both primary and highest weight, we find
\bea
U(\sigma)|\Delta,j_R,j_L;j_R,j_L\rangle &=&
e^{i\gamma \Delta+i\lambda^R_{3}j_R +i\lambda^L_{3}j_L}
\sum_{n_R=0}^{2j_R}\sum_{n_L=0}^{2j_L} (i\lambda_-^R)^{n_R}(i\lambda_-^L)^{n_L}\cr\cr
&&\qquad\qquad\times c_{n_R,n_L}e^{i\alpha\cdot P}|\Delta,j_R,j_L;j_R-n_R,j_L-n_L\rangle\cr\cr
&=&|\alpha\rangle
\eea
where
\bea
c_{n_R,n_L}&=&{1\over n_R!}{1\over n_L!}
\prod_{k_R=0}^{n_R-1}\prod_{k_L=0}^{n_L-1}
\sqrt{(2j_R-k_R)(k_R+1)(2j_L-k_L)(k_L+1)}\cr
&=&\sqrt{{2j_L\choose n_L}{2j_R\choose n_R}}
\eea

\subsection{Cost Function and Fubini-Study Metric}

As discussed for SO(2,3) in Section \ref{costone}, we could study the ${\cal F}_1$ cost function
\begin{equation}
{\cal F}_1d\sigma =|\langle\Delta,j_R,j_L;j_R,j_L|U^\dagger dU|\Delta,j_R,j_L;j_R,j_L\rangle|
\end{equation}
Again, because ${\cal F}_1$ assigns zero cost to certain gates, we choose to rather focus on the Fubini-Study
metric 
\begin{equation}
ds^2=\langle\Delta,j_R,j_L;j_R,j_L|dU^\dagger dU|\Delta,j_R,j_L;j_R,j_L\rangle 
-|\langle\Delta,j_R,j_L;j_R,j_L|U^\dagger dU|\Delta,j_R,j_L;j_R,j_L\rangle|^2
\end{equation}
as a measure of complexity.
The action of the unitary $U(\sigma)$ on the highest weight reference state $|\Delta, j_L, j_R; j_L, j_R \rangle$ again
simplifies after using the fact that the reference state is both primary and highest weight
\begin{equation}
U(\sigma ) |\Delta, j_L, j_R; j_L, j_R \rangle = e^{i\gamma \Delta } e^{i \lambda_3^L j_L } e^{i \lambda_3^R j_R } \  e^{i \alpha \cdot P} e^{i \lambda_{-}^R J_{-}^R} e^{i \lambda_{-}^L J_{-}^L} |\Delta, j_L, j_R; j_L, j_R\rangle
\end{equation}
The evaluation of the Fubini-Study metric is again a straight forward exercise, once the matrix element
\bea
\langle \Delta, j_L, j_R; j_L, j_R | U^\dag(\sigma) U(\sigma) |\Delta, j_L, j_R; j_L, j_R \rangle\label{disfour}
\eea
has been evaluated.
The operator state correspondence can again be used to relate matrix elements to correlation functions.
Matrix elements taken using highest weight states again exhibit a particularly simple dependence on the $j_L,j_R$ quantum
numbers.
By decomposing so(4) into su(2)$\times$su(2), the key formulas (\ref{sphharm1}), (\ref{bchange}) and (\ref{prodstruct}) 
apply to each su(2) factor, predicting that (\ref{disfour}) takes the form $\propto A_L^{j_L}A_R^{j_R}$.
This can be verified in specific examples.
Once again the Baker-Campbell-Hausdorff formula gives an efficient approach, for arbitrary $j_L,j_R$.
See Appendix \ref{BCH} for helpful details.
The final result is given by 
\begin{eqnarray}
&& \langle \Delta, j_L, j_R; j_L, j_R | U^\dag(\sigma) U(\sigma) |\Delta, j_L, j_R; j_L, j_R \rangle \cr\cr
&&\qquad\qquad = e^{i \Delta(\gamma - \gamma^*) + i j_L(\lambda_3^L - (\lambda_3^L)^*) + i j_R(\lambda_3^R - (\lambda_3^R)^*)  } (1 - 2 \alpha\cdot \alpha^* + \alpha \cdot \alpha \alpha^* \cdot \alpha^*)^{-\Delta} \times \cr\cr
&&\qquad\qquad \left( \frac{ 1 + \lambda_{-}^L (\lambda_{-}^L)^*   + (\alpha^*)^{\mu} M^L_{\mu\nu} \alpha^\nu }
{\sqrt{1 - 2 \alpha \cdot \alpha^* + \alpha \cdot \alpha \alpha^* \cdot \alpha^* }}\right)^{2j_L} \left( 
\frac{1+\lambda_{-}^R (\lambda_{-}^R)^* +(\alpha^*)^{\mu} M^R_{\mu\nu}\alpha^\nu}
{\sqrt{1 - 2 \alpha \cdot \alpha^* + \alpha \cdot \alpha \alpha^* \cdot \alpha^* }}\right)^{2j_R} \cr
&&
\end{eqnarray}
where 
\begin{eqnarray}
M^L &=& -\left( \begin{array}{cccc} 1 + \lambda_{-}^L(\lambda_{-}^L)^* & -i(\lambda_{-}^L (\lambda_{-}^L)^* - 1) & -i(\lambda_{-}^L + (\lambda_{-}^L)^*  ) &  -\lambda_{-}^L + (\lambda_{-}^L)^* \\ i (\lambda_{-}^L (\lambda_{-}^L)^* - 1) & 1 + \lambda_{-}^L(\lambda_{-}^L)^* & -\lambda_{-}^L + (\lambda_{-}^L)^* & i ( \lambda_{-}^L + (\lambda_{-}^L)^* )  \\ i(\lambda_{-}^L + (\lambda_{-}^L)^*  ) & \lambda_{-}^L - (\lambda_{-}^L)^* &  1 + \lambda_{-}^L(\lambda_{-}^L)^* & -i (\lambda_{-}^L (\lambda_{-}^L)^* - 1 ) \\ \lambda_{-}^L - (\lambda_{-}^L)^* & -i(\lambda_{-}^L + (\lambda_{-}^L)^*) &  i(\lambda_{-}^L (\lambda_{-}^L)^* - 1   ) &  1 + \lambda_{-}^L(\lambda_{-}^L)^* \end{array} \right)   \nonumber \\
M^R & = & -\left( \begin{array}{cccc} 1 + \lambda_{-}^R(\lambda_{-}^R)^* & -i(\lambda_{-}^R (\lambda_{-}^R)^* - 1) & -i(\lambda_{-}^R + (\lambda_{-}^R)^*  ) &  \lambda_{-}^R - (\lambda_{-}^R)^* \\ i (\lambda_{-}^R (\lambda_{-}^R)^* - 1) & 1 + \lambda_{-}^R(\lambda_{-}^R)^* & -\lambda_{-}^R + (\lambda_{-}^R)^* & -i ( \lambda_{-}^R + (\lambda_{-}^R)^* )  \\ i(\lambda_{-}^R + (\lambda_{-}^R)^*  ) & \lambda_{-}^R - (\lambda_{-}^R)^* &  1 + \lambda_{-}^R(\lambda_{-}^R)^* & i (\lambda_{-}^R (\lambda_{-}^R)^* - 1 ) \\ -\lambda_{-}^R + (\lambda_{-}^R)^* & i(\lambda_{-}^R + (\lambda_{-}^R)^*) &  -i(\lambda_{-}^R (\lambda_{-}^R)^* - 1   ) &  1 + \lambda_{-}^R(\lambda_{-}^R)^* \end{array} \right) \nonumber
\end{eqnarray}
The matrix elements needed for an evaluation of the Fubini-Study metric can now be obtained by taking appropriate derivatives
of the above result.
The final answer for the Fubini-Study metric is
\begin{eqnarray}
& & \frac{ds^2}{d\sigma^2} \nonumber \\
& = & 2(\Delta + j_L + j_R) \left( \frac{\dot{\alpha}^* \cdot \dot{\alpha} - 2 |\dot{\alpha} \cdot \alpha|^2 }{1 - 2\alpha^* \cdot \alpha + (\alpha \cdot \alpha)(\alpha^* \cdot \alpha^*) }     + 2\frac{|\dot{\alpha} \cdot \alpha^* - (\alpha^* \cdot \alpha^*) \alpha \cdot \dot{\alpha}|^2}{(1 - 2\alpha^* \cdot \alpha + (\alpha \cdot \alpha)(\alpha^* \cdot \alpha^*))^2}\right)   \nonumber \\
& & +\sum_{A =L,R} \left( - 2 j_A \left( \frac{ (\alpha^*)^\mu M^A_{\mu\nu} \dot{\alpha}^\nu   (\dot{\alpha}^*)^\sigma M^A_{\sigma\tau} \alpha^\tau  }{\left( 1 + \lambda^A_{-} (\lambda^A_{-})^* + (\alpha^*)^\mu M^A_{\mu\nu} \alpha^\nu  \right)^2} + \frac{(\dot{\alpha}^* )^\mu M^A_{\mu\nu} \dot{\alpha}^\nu  }{\left( 1 + \lambda^A_{-} (\lambda^A_{-})^* + (\alpha^*)^\mu M^A_{\mu\nu} \alpha^\nu  \right)}     \right) \right. \nonumber \\
& & + 2 j_A \frac{1 - 2\alpha^* \cdot \alpha + (\alpha \cdot \alpha)(\alpha^* \cdot \alpha^*)}{\left( 1 + \lambda^A_{-} (\lambda^A_{-})^* + (\alpha^*)^\mu M^A_{\mu\nu} \alpha^\nu  \right)^2 } \dot{\lambda}^A_{-} (\dot{\lambda}_{-}^A)^* \nonumber \\
& & \left. +2 i j_A \frac{ \left(  (\alpha^*)^\mu L^A_{\mu\nu} \dot{\alpha}^\nu  - (\alpha^* \cdot \alpha^* ) 
\alpha^\mu L^A_{\mu\nu} \dot{\alpha}^\nu \right) (\dot{\lambda}_{-}^A)^{*}  
+ \left(  \alpha^\mu L^{A \dag}_{\mu\nu}(\dot{\alpha}^*)^\nu-(\alpha \cdot \alpha)(\alpha^*)^\mu L^{A\dag}_{\mu\nu} (\dot{\alpha}^* )^\nu \right) \dot{\lambda}_{-}^A}{\left( 1 + \lambda^A_{-} (\lambda_{-}^A)^* + (\alpha^*)^\mu M^A_{\mu\nu} \alpha^\nu  \right)^2} \right) \nonumber \\
& & \label{FSMetricSO42}
\end{eqnarray}
where
\bea
L^L &\equiv& -i (1 + \lambda^L_{-} (\lambda^L_{-})^* )^2 \frac{\partial}{\partial (\lambda_{-}^L)^* } 
{M^L\over 1 + \lambda^L_{-} (\lambda_{-}^L)^* }\cr\cr\cr
&=& \left( \begin{array}{cccc} 0 & 2 \lambda_{-}^L & 1 - (\lambda_{-}^L)^2 & i(1 + (\lambda_{-}^L)^2 ) \\
-2 \lambda_{-}^L & 0 & i(1 + (\lambda_{-}^L)^2  ) & -1 + (\lambda_{-}^L)^2 \\ 
-1 + (\lambda_{-}^L)^2 & - i(1 + (\lambda_{-}^L)^2 ) & 0 & 2\lambda_{-}^L  \\ -i(1 + (\lambda_{-}^L)^2 ) & (1 - (\lambda_{-}^L)^2 ) & - 2\lambda_{-}^L & 0 \end{array} \right)
\eea
\bea
L^R &\equiv& -i (1 + \lambda^R_{-} (\lambda^R_{-})^* )^2 \frac{\partial}{\partial (\lambda_{-}^R)^* }
{ M^R\over 1 + \lambda^R_{-} (\lambda_{-}^R)^* }\cr\cr\cr
& = & \left( \begin{array}{cccc} 0 & 2 \lambda_{-}^R & 1 - (\lambda_{-}^R)^2 & -i(1 + (\lambda_{-}^R)^2 ) \\
-2 \lambda_{-}^R & 0 & i(1 + (\lambda_{-}^R)^2  ) & 1 - (\lambda_{-}^R)^2 \\ 
-1 + (\lambda_{-}^R)^2 & - i(1 + (\lambda_{-}^R)^2 ) & 0 & -2\lambda_{-}^R  \\ i(1 + (\lambda_{-}^R)^2 ) & (-1 + (\lambda_{-}^R)^2 ) & 2\lambda_{-}^R & 0 \end{array} \right)
\eea

For a scalar primary, all dependence on $\lambda_-^R$ and $\lambda_-^L$ drops out and we obtain a metric defined on an 
8 dimensional space with complex coordinates $\alpha^\mu$.
This space can be identified with the coset space SO(2,4)/SO(2)$\times$SO(4) \cite{Chagnet:2021uvi,Gibbons:1999rb},
where we recognize the stability group of a scalar primary which is SO(2)$\times$SO(4).
The stability group of a spinning primary is SO(2)$\times$SO(2)$\times$SO(2).
In this case the metric is defined on an 12 dimensional space with complex coordinates $\alpha^\mu$, $\lambda^R_-$
and $\lambda_-^L$.
This space is the coset space SO(2,4)/SO(2)$\times$SO(2)$\times$SO(2).

\subsection{Fubini-Study metric from coadjoint orbit}

The result (\ref{FSMetricSO42}) is again in complete agreement with the metric obtained by studying the geometry
of coadjoint orbits of SO(2,4).
The computation is a simple generalization of that of Section \ref{coadjoint}, so we need only sketch the main points. 
Working in the defining representation and after imposing unitarity we find
\bea
U = e^{i \alpha \cdot P} e^{i \sigma \cdot K} e^{ i\gamma_R D +  \log\left( \gamma \right) D} \prod_{A = L,R} e^{i \lambda_{-}^A J_{-}^A} e^{i \lambda_{+}^A J_{+}^A} e^{ i \lambda_{3 R}^A J_3^A + \left( \log(\gamma) -\log(\Lambda^A) \right) J_3^A}
\eea
where $\gamma$ is defined in \eqref{defgam}, 
\bea
\sigma_{u}  =  -\frac{\partial}{\partial \alpha^{\mu}} \log\left( \gamma \right) \qquad
 \lambda_{+}^A  = \frac{\partial}{\partial \lambda^A_{-}} \log\left( \Lambda^A \right) 
\eea
and 
\bea
\Lambda^A & = & -(1 + \lambda_{-}^A (\lambda_{-}^A)^* ) - (\alpha^*)^\mu M_{\mu\nu}^A \alpha^\nu \nonumber
\eea
Notice that after imposing unitarity $U(\sigma)$ depends on the complex vector $\alpha^\mu$ with four components, the 
two complex numbers $\lambda_-^R,\lambda_-^L$ and 
the three real numbers $\gamma_R,\lambda_{3R}^R,\lambda_{3R}^L$.
This gives a total of 15 real parameters, which matches the dimension of so(2,4).
The computation of the Maurer-Cartan form, the symplectic form and the metric on the coadjoint orbit is now straightforward.
To evaluate the symplectic form we choose
\begin{equation}
\eta = i \Delta D + i \frac{i j_L}{2}\left( -L_{34}  - L_{12} \right) + i \frac{j_R}{2 i}\left( -L_{34}  + L_{12} \right)
\end{equation}
Notice that the stabilizer algebra of $\eta$ is so(2)$\times$so(2)$\times$so(2) which matches the stabilizer algebra of the
spinning primary reference state.
The resulting metric on the coadjoint orbit is in complete with the Fubini-Study metric (\ref{FSMetricSO42}).

\section{Discussion and Conclusions}\label{conclusions}

By studying quantum circuits that are irreducible representations of the Lorentzian conformal groups so(2,3) and so(2,4),
we have obtained explicit formulas for the Fubini-Study metrics that give measures of computational complexity.
In particular, by studying circuits that start from arbitrary spinning primary states we have generalized the formulas obtained
in \cite{Chagnet:2021uvi}, which considered scalar primary states.
The metrics we have obtained can be reproduced, in complete detail, from the geometry of coadjoint orbits.

Our results show a remarkable simplicity that warrants further discussion.
The spinning primary reference state is labeled by it's quantum numbers with respect to the Cartan subalgebra of the
conformal group.
This is a dimension $\Delta$ and a single spin $j$ for so(2,3) and two spins $(j_L,j_R)$ for so(2,4).
We have seen that the Fubini-Study metric is easily computed once the overlap
\bea
I=\langle\phi |U^\dagger (\sigma)U(\sigma)|\phi\rangle
\eea
has been evaluated. 
Setting $I$ to 1 gives a set of unitarity conditions.
Before imposing unitarity, taking derivatives of $I$ we can evaluate the matrix elements that appear in the Fubini-Study metric.
Our results show that $I$ has a remarkably simple structure.
Up to some overall phases which are trivially determined, there is a factor for each quantum number of the Cartan subalgebra.
In three dimensions we have
\bea
I= (A_\Delta)^\Delta (A)^j
\eea
while in four dimensions we have
\bea
I= (A_\Delta)^\Delta (A_L)^{j_L} (A_R)^{j_R}\label{frstne}
\eea
For both formulas we have
\bea
A_\Delta={1\over 1-2\alpha\cdot\alpha^*+|\alpha\cdot\alpha|^2}\label{scndne}
\eea
The spin factors $A$ in \eqref{frstne} and $A_L,A_R$ in \eqref{scndne} also have a universal form.
Collectively denoting these factors as $A_{\rm spin}$ we have
\begin{equation}
A_{\rm spin}=\frac{1+\lambda_{-}\lambda_{-}^*+\alpha^*\cdot M\cdot\alpha}
{\sqrt{1-2\alpha\cdot\alpha^*+|\alpha \cdot \alpha|^2}}\label{thrdne}
\end{equation}
In the above formula, $M$ is a $d\times d$ matrix for so(2,$d$).
Our results suggest that this matrix again has a universal form, easily written down using the defining representation of
so($d$).
For example, for so(2,3) we use the defining representation of so(3) which gives matrices
\begin{eqnarray}
J_3=\left[\begin{array}{ccc} 0 & -i & 0 \\ i & 0 & 0 \\ 0& 0 & 0   \end{array}\right] \qquad
J_+ = \left[\begin{array}{ccc} 0 & 0 & -1 \\ 0 & 0 & -i \\ 1& i & 0   \end{array}\right] \qquad
J_- = J_-^\dagger
\end{eqnarray}
In terms of these matrices $M$ can be written as
\bea
M &=&(1-\lambda_-\lambda_-^*)J_3-i\lambda_-^* J_++i\lambda_-J_--(1+\lambda_-\lambda_-^*)I_3\label{Mfirst}
\eea
where $I_3$ is the three dimensional identity matrix.
To illustrate why we claim $M$ takes a universal form, consider so(2,4) where we use the defining representation of so(4).
The relevant matrices are
\bea
J_{3}^L &=& \frac{1}{2i}(L_{12} + L_{34}) 
=\frac{1}{2}\left[
\begin{array}{cccc} 0 & -i & 0 & 0 \\ i & 0 & 0 & 0  \\ 0 & 0 & 0 & -i \\ 0 & 0 & i & 0\end{array}
\right]\cr\cr\cr
J_{+}^L &=& \frac{1}{2}( -L_{13} - i L_{14} - i L_{23} + L_{24}    ) 
= \frac{1}{2}\left[
\begin{array}{cccc} 0 & 0 & -1 & -i \\ 0 & 0 & -i & 1  \\ 1 & i & 0 & 0 \\ i & -1 & 0 & 0   \end{array} 
\right] 
\qquad J_{-}^L = (J_+^L)^\dagger
\eea
\bea
J_{3}^R & = & \frac{i}{2} (L_{34} - L_{12}) 
= \frac{1}{2}\left[
\begin{array}{cccc} 0 & -i & 0 & 0 \\ i & 0 & 0 & 0  \\ 0 & 0 & 0 & i \\ 0 & 0 & -i & 0   \end{array} 
\right]\cr\cr\cr 
J_{+}^R &=&  \frac{1}{2}(-L_{13} + i L_{14} - i L_{23} - L_{24})  
=\frac{1}{2}\left[
\begin{array}{cccc} 0 & 0 & -1 & i \\ 0 & 0 & -i & -1  \\ 1 & i & 0 & 0 \\ -i & 1 & 0 & 0 \end{array} 
\right] 
\qquad 
J_{-}^L = (J_{+}^R)^\dagger
\eea
Using these matrices we can write
\bea
M^L&=&2\big((1-\lambda^L_-(\lambda_-^L)^*)J_3^L - i (\lambda_-^L)^*J_+^L+i\lambda_-^L J_-^L \big) 
-(1+\lambda^L_-(\lambda^L_-)^*) I_4 \label{MSecond}
\eea
and
\begin{eqnarray}
M^R&=&2\big((1-\lambda^R_-(\lambda_-^R)^*)J_3^R-i(\lambda_-^R )^* J_+^R+i\lambda_-^R J_-^R \big) 
-(1+\lambda^R_-(\lambda^R_-)^*) I_4\label{MThird}
\end{eqnarray}
where $I_4$ is the four dimensional identity matrix.
Notice that \eqref{Mfirst}, \eqref{MSecond} and \eqref{MThird} are almost in complete agreement.
The only difference is in the coefficient of the term linear in the Lie algebra generators.
This coefficient will depend on the choice of normalization for the Lie algebra elements.
In the current setting notice that $J_3$ has integer eigenvalues while $J_3^L$ and $J_3^R$ both have half integer
eigenvalues.

Based on our experience with SO(2,3) and SO(2,4) we might try to guess a formula for the overlap $I$ computed in 
$d$ dimensions, where the relevant conformal group is SO(2,$d$).
The overlap has taken a striking form with a factor for each element in the Cartan sub algebra.
We expect that the factor $A_\Delta$ will appear in the general SO(2,$d$) case, motivated by the answer from circuits
which are in a representation of SO(2,$d$) and start from a scalar primary.
In contrast to this, we expect that the factors associated with the spin, denoted $A_{\rm spin}$ above, will change.
It is because the compact subgroup associated to spin is SO(3)$\simeq$SU(2) for SO(2,3), and is SO(4)=SU(2)$\times$SU(2)
for SO(2,4) that the $A_{\rm spin}$ factors have taken such a uniform form in our results.

The Fubini-Study metrics we have computed furnish a concrete description of the spaces relevant for the computation
of complexity in the conformal field theory.
Using these metrics we can study geodesics on the space of unitaries which correspond to trajectories emanating 
from a spinning primary.
The time evolution resulting from the conformal field theory dynamics will also produce a trajectory on this same space,
and for short enough times we expect the geodesic route to match the time evolution trajectory.
This produces a linear growth in complexity for short time scales.
However, we expect that at late enough times this growth stops and complexity saturates \cite{Susskind:2018pmk}.
This saturation has been related to the appearance of conjugate points, a global feature of the space of 
geodesics \cite{Balasubramanian:2021mxo}.
Conjugate points appear where different geodesics reconnect, so that they reflect the fact that there are shorter paths 
that intersect the time evolution trajectory. 
These shorter paths become the geodesics relevant for the computation of complexity, and the linear growth in
complexity halts.
Do our geometries exhibit conjugate points?
The basic theorem from Riemannian geometry, governing the appearance of conjugate points, is stated in
terms of Jacobi fields \cite{peterson}.
Recall that, for a given geodesic $\gamma$ any field $X$ which obeys
\bea
\nabla_{\dot\gamma}\nabla_{\dot\gamma} X+R(\dot\gamma,X)\dot\gamma=0
\eea
is called a Jacobi field.
The basic theorem relating  Jacobi fields to conjugate points states \cite{peterson}

\noindent
{\bf Theorem} Let $M$ be a Riemannian manifold and let $\gamma:[a,b]\to M$ be a geodesic. 
Then $q=\gamma(t_0)$ is a conjugate point of $p=\gamma(a)$ if and only if there exists a non-vanishing 
Jacobi field $X$ along $\gamma$ so that $X(a)=0$ and $X(t_0)=0$.

In general the construction of Jacobi fields has to be tackled numerically.
An important special case, which can be handled analytically is when the manifold has constant sectional curvature.
Let $\sigma_p$ denote the two-dimensional linear subspace of the tangent space at a point $p$ of the manifold.
The sectional curvature $K(\sigma_p)$ is defined geometrically as the Gaussian curvature of the surface which has the plane 
$\sigma_p$ as a tangent plane at $p$.
A manifold with constant section curvature has the same sectional curvature at all points $p$.
If a manifold has constant sectional curvature $k$ we know that it is an Einstein manifold and that it's Riemann tensor 
is given by the Kulkarni-Nomizu product of the metric with itself
\bea
R={k\over 2}g\KN g
\eea
Recall that the metric obtained from scalar primaries has constant negative sectional curvature.
In this case the Jacobi fields can be constructed explicitly and it is known that these spaces, have no conjugate 
points \cite{helgason} as pointed out in \cite{Chagnet:2021uvi}.
The geometries we have found using a spinning reference state have constant scalar curvature, but they
are not Einstein manifolds and do not have constant sectional curvature. 
We have not considered the interesting problem of constructing the Jacobi fields associated to our complexity manifolds, and
the subsequent implications for possible conjugate points.
Constructing the Jacobi fields would allow us to rule out the existence of conjugate points, or to explicitly construct them,
signaling the saturation of complexity.
Towards this end it is useful to note that even for the case of the spinning reference states the metrics we have
constructed are K\"ahler
\begin{equation}
g_{\mu \bar{\nu}} = \partial_\mu \partial_{\bar{\nu}} V_K.  
\end{equation}
with K\"ahler potential 
\begin{equation}
V_K = \log\left( (1 - 2 \alpha \cdot \alpha^* + \alpha \cdot \alpha \alpha^* \cdot \alpha^*)^{-(\Delta + j)} \Lambda^{2j} \right)   \label{KPot}
\end{equation}
In this case, the Ricci tensor can be written as
\begin{equation}
R_{\mu\bar{\nu}} = -\frac{1}{2}\partial_\mu \partial_{\bar{\nu}} \log\left( \det g \right)
\end{equation}

There is one clear deficiency with our analysis: our allowed gates have been restricted to conformal transformations.
Since conformal transformations are symmetries, we have good motivation to assign allowed gates the same
computational cost, and hence the arbitrariness associated with choosing a cost function is removed.
However, this also means that starting with any reference state in a given conformal multiplet, our circuit can never
produce a target state that belongs to a different multiplet.
Symmetry operations do not provide a universal set of gates from which any desired circuit can be synthesized.
In fact, if the set of allowed gates is not large enough we might not even see a saturation of complexity, because with the
limited set of gates at our disposal, we might not have access to the more efficient unitaries which are responsible for the
saturation of complexity.
A first step towards a more complete set of gates might involve studying theories with a higher spin symmetry and allowing
gates that correspond to the higher spin symmetries.
We stress however that it is only after a complete set of allowed gates is defined, that the study of the saturation of 
complexity can be performed.
The problem of characterizing a suitable set of allowed gates seems difficult.
One possibility would be to take the minimal set of gates with which field theory computations reproduce expectations 
derived from the holographic dual theory.
Clearly there are still very basic open questions that remain unresolved.

\section*{Acknowledgement}
This work is supported by a Simons Foundation Grant Award ID 509116 and by the South African Research Chairs initiative 
of the Department of Science and Technology and the National Research Foundation.
During the final stages of this project MK was also
supported by the Basic Science Research Program (2020R1A6A1A03047877) of the National
Research Foundation of Korea funded by the Ministry of Education through Center for Quantum Spacetime 
(CQUeST) of Sogang University.

\begin{appendix}

\section{Baker-Campbell-Hausdorff Identites}\label{BCH}

We will derive an identity for the operator $e^{-i\alpha^*\cdot K}e^{i\alpha\cdot P}$ in $d=3$ Euclidean space.
An su(2)$\times$su(2) subgroup of so(2,3) plays an important role.
The generators of the subgroup are defined by
\bea
J_+^L=v_+\cdot K\qquad J_-^L=-v_-\cdot P\qquad J_3^L=-{1\over 2}D + v_{+\mu}v_{-\nu}L_{\mu\nu}
\label{firstsu}
\eea
and
\bea
J_+^R=-v_+\cdot P\qquad J_-^R=v_-\cdot K\qquad J_3^R={1\over 2}D + v_{+\mu}v_{-\nu}L_{\mu\nu}
\label{secondsu}
\eea
where $v_+\cdot v_+=0=v_-\cdot v_-$ and $v_+\cdot v_-$.
To obtain a null vector in Euclidean space the vectors $v_{+\mu}$ and $v_{-\nu}$ must have complex components.
We can write $\alpha_\mu$ and $\alpha^*_\mu$  in terms of two null vectors as follows
\bea
\alpha_\mu &=& i\sqrt{\alpha\cdot\alpha}(v_{+\mu}-v_{-\nu})\cr\cr
\alpha^*_\mu&=&-{i\over\sqrt{\alpha\cdot\alpha}}(C_-v_{+\mu}-C_+v_{-\nu})\cr\cr
C_\pm &=& -\alpha\cdot\alpha^*\pm i\sqrt{\alpha\cdot\alpha\alpha^*\cdot\alpha^*-(\alpha\cdot\alpha^*)^2}
\eea
We can also express
\bea
v_{+\mu}&=&{1\over 2C\sqrt{\alpha\cdot\alpha}}(C_+\alpha_{\mu}+\alpha\cdot\alpha \alpha^*_\mu)
\cr\cr
v_{-\mu}&=&{1\over 2C\sqrt{\alpha\cdot\alpha}}(C_-\alpha_{\mu}+\alpha\cdot\alpha \alpha^*_\mu)
\eea
where $C={i\over 2}(C_+-C_-)$.
Noting that
\bea
i\alpha\cdot P=\sqrt{\alpha\cdot\alpha}(J_+^R-J_-^L)\qquad
-i\alpha^*\cdot K={1\over\sqrt{\alpha\cdot\alpha}}(C_+J_-^R-C_-J_+^L)
\eea
we can use the Baker-Campbell-Hausdorff identity
\bea
e^{a_+^LJ_+^L+a_-^RJ_-^R}e^{b_-^L J_-^L+b_+^RJ_+^R}
&=&e^{{b_-^L\over 1+a_+^Lb_-^L}J_-^L+{b_+^R\over 1+a_-^Rb_+^R}J_+^R}
e^{2\log (1+a_+^Lb_-^L)J_3^L-2\log (1+a_-^Rb_+^R)J_3^R}\cr\cr
&\times&e^{{a_+^L\over 1+a_+^Lb_-^L}J_+^L+{b_-^R\over 1+a_-^Rb_+^R}J_-^R}\label{BCH1}
\eea
to obtain
\bea
e^{-i\alpha^*\cdot K}e^{i\alpha\cdot P}=e^{g\cdot P}
e^{-\log (1+C_++C_-+C_+C_-)D+\log\left({1+C_-\over 1+C_+}\right)v_{+\mu}v_{-\nu}L_{\mu\nu}}
e^{h\cdot K}
\eea
where the exact expressions for the vectors $g_\mu$ and $h_\mu$ are easy to obtain but will not be needed.

Next, we will derive an identity for the operator $e^{-i\lambda_-^* J_+}e^{g\cdot P}e^{i\lambda_-^* J_+}$,
again in $d=3$ Euclidean space. 
Using the result (\ref{BCH1}) as well as
\bea
e^{-i\lambda_-^* J_+}e^{g\cdot P}e^{i\lambda_-^* J_+}=e^{\tilde{g}\cdot P}\qquad\qquad
e^{i\lambda_-J_-}e^{h\cdot K}e^{-i\lambda_-J_-}=e^{\tilde{h}\cdot K}
\eea
the matrix element
\bea
\langle \Delta,j;j|e^{-i\lambda_-^* J_+}e^{-i\alpha^*\cdot K}e^{i\alpha\cdot P}e^{i\lambda_-J_-}|\Delta,j;j\rangle
\eea
simplifies to
\bea
e^{-\log (1+C_++C_-+C_+C_-)\Delta}
\langle \Delta,j;j|e^{-i\lambda_-^* J_+}e^{\log\left({1+C_-\over 1+C_+}\right)v_{+\mu}v_{-\nu}L_{\mu\nu}}e^{i\alpha\cdot P}e^{i\lambda_-J_-}|\Delta,j;j\rangle
\eea
If we now write $v_{+\mu}v_{-\nu}L_{\mu\nu}=j_3 J_3+j_-J_-j_+J_+$, then by making use of the following
identity (again derived using the Baker-Campbell-Hausdorff formula)
\bea
e^{-iJ_+\lambda_-^*}e^{j_3 J_3+j_-J_-j_+J_+}e^{iJ_-\lambda_-}= e^{J_- A}e^{2 J_3 B}e^{J_+C}\label{BCH2}
\eea
where
\bea
A&=&\frac{\left(2 j_- \left(e^{|\vec{j}|}-1\right)+i \lambda_- \left(j_3 \left(-e^{|\vec{j}|}\right)+\left(e^{|\vec{j}|}+1\right) |\vec{j}|+j_3\right)\right)}{d_A}\cr\cr
d_A&=&\lambda_- \lambda_-^* |\vec{j}|-j_3 (\lambda_- \lambda_-^*-1) \left(e^{|\vec{j}|}-1\right)+e^{|\vec{j}|} \left((\lambda_- \lambda_-^*+1) |\vec{j}|-2 i j_- \lambda_-^*+2 i j_+ \lambda_-\right)\cr\cr
&&\qquad +|\vec{j}|+2 i j_- \lambda_-^*-2 i j_+ \lambda_-\cr\cr
|\vec{j}|&\equiv& \sqrt{j_3^2+4 j_-j_+}\cr\cr
B&=&\log \left(\frac{e^{-\frac{1}{2} |\vec{j}|} d_A}{2 |\vec{j}|}\right)\cr\cr
C&=&\frac{\left(2 j_+ \left(e^{|\vec{j}|}-1\right)-i \lambda_-^* \left(j_3 \left(-e^{|\vec{j}|}\right)+\left(e^{|\vec{j}|}+1\right) |\vec{j}|+j_3\right)\right)}{d_A}
\eea
we obtain (\ref{GenRes}).

A few comments are in order.
First, notice that the su(2)$\times$su(2) subgroup identified above can be defined for any so(2,$d$) with $d\ge 2$.
In addition, the formulas \eqref{BCH1} and \eqref{BCH2} were established using a specific su(2) representation.
It is however simple to check, by varying the representation used, that the identities are independent of the specific
representation used.

The Baker-Campbell-Hausdorff formula is also useful for the $d=4$ computation of complexity.
Using the logic above in \eqref{firstsu} and \eqref{secondsu} to extract an su(2)$\times$su(2) subgroup and
imposing that $U(\sigma)$ is unitary, we find
\begin{eqnarray}
&&\langle \Delta, j_L, j_R; j_L, j_R | U^\dag(\sigma) U(\sigma) |\Delta, j_L, j_R; j_L, j_R \rangle
= e^{i \Delta(\gamma - \gamma^*) + i j_L(\lambda_3^L - (\lambda_3^L)^*) + i j_R(\lambda_3^R - (\lambda_3^R)^*)}
\cr\cr
&&\qquad\times  (1 - 2 \alpha\cdot \alpha^* + \alpha \cdot \alpha \alpha^* \cdot \alpha^*)^{-\Delta}
\langle \Delta, j_L, j_R; j_L, j_R | e^{-i (\lambda_{-}^L)^* J_{+}^L } e^{-i (\lambda_{-}^R)^* J_{+}^R } \cr\cr
&&\qquad\times e^{i \log\left( \frac{1 + C_{-}}{1 + C_{+} }\right) \frac{\alpha_\mu \alpha_\nu^* L^{\mu\nu}}{\sqrt{(\alpha\cdot \alpha)(\alpha^* \cdot \alpha^*) - (\alpha\cdot \alpha^*)^2 }}  }   e^{i \lambda_{-}^R J_{-}^R} e^{i \lambda_{-}^L J_{-}^L}  |\Delta, j_L, j_R; j_L, j_R \rangle  \nonumber
\end{eqnarray}
The final matrix element takes the form of a product of two copies of SU(2) matrix elements.
To evaluate the matrix element above, apply two copies of the SU(2) Baker-Campbell-Hausdorff formula.
The following decomposition shows how to translate between so(4) and su(2)$\times$su(2), which is needed
to carry the computation out
\begin{eqnarray}
L_{12} & = & i\left( J_3^L + J_3^R \right)  \nonumber \\
L_{13} & = & \frac{1}{2} \left( J_{-}^L + J _{-}^R - J_{+}^L - J_{+}^R \right) \nonumber \\
L_{14} & = & -\frac{i}{2}\left( -J_{-}^L + J_{-}^R - J_{+}^L + J_{+}^R \right) \nonumber \\
L_{23} & = & \frac{i}{2}\left( J_{-}^L + J_{-}^R + J_{+}^L + J_{+}^R  \right) \nonumber \\
L_{24} & = & \frac{1}{2}\left( -J_{-}^L + J_{-}^R + J_{+}^L - J_{+}^R \right) \nonumber \\
L_{34} & = & i \left( J_{3}^L - J_{3}^R\right)
\end{eqnarray}

\section{More Geometry}\label{MG}

The computation of complexity has been reduced to evaluating the length of a geodesic, with the starting point of
the geodesic determined by a choice of reference state and ending point by a choice of target state.  
Our focus in this Appendix is on the dependence of complexity on the choice of reference state.
In the discussion that follows, we focus on SO(2,3).  
Choose the reference state $|\phi_0\rangle = |\Delta,j; j-n\rangle$ which has definite scaling dimension $\Delta$ and 
$L_3$ spin $j-n$.  
The resulting eight dimensional Fubini-Study metric is
\begin{eqnarray}
ds^2
&=&2(\Delta + j - n) \left( 
\frac{d\alpha^* \cdot d\alpha - 2 |d\alpha \cdot \alpha|^2 }{1 - 2\alpha^* \cdot \alpha + (\alpha \cdot \alpha)(\alpha^* \cdot \alpha^*) }     
+ 2\frac{|d\alpha\cdot \alpha^* - (\alpha^* \cdot \alpha^*) \alpha \cdot d\alpha|^2}{(1 - 2\alpha^* \cdot \alpha + (\alpha \cdot \alpha)(\alpha^* \cdot \alpha^*))^2}\right)   \nonumber \\
& & - 2 (j-n) \left( \frac{ (\alpha^*)^\mu M_{\mu\nu} d\alpha^\nu   (d\alpha^*)^\sigma M_{\sigma\tau} \alpha^\tau  }{\left( 1 + \lambda_{-} \lambda_{-}^* 
+ (\alpha^*)^\mu M_{\mu\nu} \alpha^\nu  \right)^2} + \frac{(d\alpha^* )^\mu M_{\mu\nu}d\alpha^\nu  }
{\left( 1 + \lambda_{-} \lambda_{-}^* + (\alpha^*)^\mu M_{\mu\nu} \alpha^\nu  \right)}     \right) \nonumber \\
& & + 2(j-n) \frac{\gamma^2}{\Lambda^2} 
(d\alpha \cdot L \cdot \sigma) (d\alpha^* \cdot L^\dag \cdot \sigma^*) \nonumber \\
& & + 2( j - n(n + 1 - 2j)) \frac{\gamma^2}{\Lambda^2} (-i d\lambda_{-} - d\alpha\cdot L \cdot \sigma)(id\lambda^*_{-} + d\alpha\cdot L^\dag \cdot \sigma^* )   \label{GenJ3Metric}
\end{eqnarray}
where $\Lambda$ is defined in \eqref{forp}, $M$ in \eqref{deffM} and $L$ in \eqref{deffL}.
The scalar curvature of this geometry is
\begin{equation}
R = \frac{-2\Delta^3-6 j (1+j) (j-n)^2+6 (j-n)^4+18\Delta^2 (j+2 j n-n^2)+2 \Delta (2 (j-n)^2+(j+2 j n-n^2)^2)}{(j + j^2 - (n-j)^2)\Delta ( (n-j)^2 - \Delta^2  )}
\end{equation}
which is invariant under $n \rightarrow 2j-n$ as expected.
It is noteworthy that the scalar curvature has a non-trivial dependence on $n$ and hence, on the initial reference state.
 
There are many initial states that could be considered which, for example, need not be eigenstates of $L_3$.
As a nice way to summarize this more general set up, we have been able to write the Fubini-Study metric for
a reference state of scaling dimension $\Delta$ and arbitrary spin as
\begin{eqnarray}
\frac{ds^2}{d\sigma^2}
&=& 2(\Delta + \langle J_{3} \rangle ) \left( \frac{\dot{\alpha}^* \cdot \dot{\alpha} - 2 |\dot{\alpha} \cdot \alpha|^2 }{1 - 2\alpha^* \cdot \alpha + (\alpha \cdot \alpha)(\alpha^* \cdot \alpha^*) }     + 2\frac{|\dot{\alpha} \cdot \alpha^* - (\alpha^* \cdot \alpha^*) \alpha \cdot \dot{\alpha}|^2}{(1 - 2\alpha^* \cdot \alpha + (\alpha \cdot \alpha)(\alpha^* \cdot \alpha^*))^2}\right)   \nonumber \\
& & - 2 \langle J_{3} \rangle \left( \frac{ (\alpha^*)^\mu M_{\mu\nu} \dot{\alpha}^\nu   (\dot{\alpha}^*)^\sigma M_{\sigma\tau} \alpha^\tau  }{\left( 1 + \lambda_{-} \lambda_{-}^* + (\alpha^*)^\mu M_{\mu\nu} \alpha^\nu  \right)^2} + \frac{(\dot{\alpha}^* )^\mu M_{\mu\nu} \dot{\alpha}^\nu  }{\left( 1 + \lambda_{-} \lambda_{-}^* + (\alpha^*)^\mu M_{\mu\nu} \alpha^\nu  \right)}     \right) \nonumber \\
& & + 2\langle J_{3} \rangle \frac{\gamma^2}{\Lambda^2} ( \dot{\alpha} \cdot L \cdot \sigma) (\dot{\alpha}^* \cdot L^\dag \cdot \sigma^*) -  \frac{4 \langle   -L_{31}  - i L_{23}\rangle}{\gamma^3 \Lambda^2} \left( e^{i \lambda_{3R} }  d\alpha^* \cdot \tilde{u} \cdot L \cdot d\alpha \right)\nonumber \\ 
& &   + \frac{4 \langle   L_{31}  - i L_{23}\rangle}{{\gamma^3 \Lambda^2}} \left( e^{-i \lambda_{3R} }  d\alpha \cdot u \cdot L^\dag \cdot d\alpha^* \right) 
-\left(  \langle J_{+} J_{-} + J_{-} J_{+}\rangle  -  2\langle J_{+}\rangle \langle J_{-}\rangle \right) f_{+} f_{-} \nonumber \\
& & -  \left(  \langle J_{-} J_{-} \rangle  -  \langle J_{-}\rangle^2  \right) f_{-}^2  - \left(  \langle J_{+} J_{+} \rangle  -  \langle J_{+}\rangle^2 \right) f_{+}^2 
- \left(  \langle J_{3} J_{3} \rangle  -  \langle J_{3}\rangle^2 \rangle \right) f_{3}^2\nonumber \\
& &  - \left(  \langle J_{3} J_{+} + J_{+} J_{3}\rangle  -  2\langle J_{3}\rangle \langle J_{+}\rangle \right) f_{3} f_{+} 
- \left(  \langle J_{3} J_{-} + J_{-} J_{3}\rangle  -  2\langle J_{3}\rangle \langle J_{-}\rangle \right) f_{3} f_{-} 
   \label{GenSpinMetric}
\end{eqnarray} 
where $\langle \cdot \rangle$ refers to the expectation value w.r.t. the initial state and 
\begin{eqnarray}
u_{ij} &=& (1 - \delta_{ij}) \left( -2(\alpha_i \alpha^*_j    +   \alpha_j \alpha^*_i ) - 2 \alpha_i \alpha_j (\alpha^* \cdot \alpha^* - 2 (\alpha^*_{i})^2 - 2 (\alpha^*_j)^2  )    - 2 \alpha^*_i \alpha^*_j (\alpha \cdot \alpha - 2 (\alpha_{i})^2 - 2 (\alpha_j)^2  ) \right. \nonumber \\
& & \left.  +  4 (\alpha_i \alpha^*_j + \alpha_j \alpha^*_i)(\alpha\cdot \alpha^* - \alpha_i \alpha^*_i - \alpha_j \alpha^*_j)  + 4 \alpha_i \alpha^*_j (1 - \alpha \cdot \alpha^*) \right) \nonumber \\
& & + \delta_{ij} \left( 1 - 2 \alpha \cdot \alpha^*  + 4 \alpha_i \alpha^*_i (\alpha \cdot \alpha^* - \alpha_i \alpha^*_i ) + (\alpha \cdot \alpha - 2 \alpha_i \alpha_i) (\alpha^* \cdot \alpha^* - 2 \alpha^*_i \alpha^*_i)\right) \nonumber \\
\tilde{u}_{ij}
&=& (1 - \delta_{ij}) \left( -2(\alpha_i \alpha^*_j    +   \alpha_j \alpha^*_i ) - 2 \alpha_i \alpha_j (\alpha^* \cdot \alpha^* - 2 (\alpha^*_{i})^2 - 2 (\alpha^*_j)^2  )    - 2 \alpha^*_i \alpha^*_j (\alpha \cdot \alpha - 2 (\alpha_{i})^2 - 2 (\alpha_j)^2  ) \right. \nonumber \\
& & \left.  +  4 (\alpha_i \alpha^*_j + \alpha_j \alpha^*_i)(\alpha\cdot \alpha^* - \alpha_i \alpha^*_i - \alpha_j \alpha^*_j)  + 4 \alpha_j \alpha^*_i (1 - \alpha \cdot \alpha^*) \right) \nonumber \\
& & + \delta_{ij} \left( 1 - 2 \alpha \cdot \alpha^*  + 4 \alpha_i \alpha^*_i (\alpha \cdot \alpha^* - \alpha_i \alpha^*_i ) + (\alpha \cdot \alpha - 2 \alpha_i \alpha_i) (\alpha^* \cdot \alpha^* - 2 \alpha^*_i \alpha^*_i)\right)\nonumber\\
f_{+} &=&  i \frac{\gamma}{\Lambda} \left( \dot{\lambda}_{-}^* - i \dot{\alpha}^* \cdot L^\dag \cdot \sigma^*    \right)e^{-i \lambda_{3R}} \nonumber \\
f_{-}&=&i \frac{\gamma}{\Lambda} \left(\dot{\lambda}_{-} - i \dot{\alpha} \cdot L \cdot \sigma    \right) e^{i\lambda_{3R} }\nonumber\\
f_3 &=&\frac{\gamma}{\Lambda}\left( 
\left(\dot\alpha\cdot{\partial\over\partial \alpha}+\dot{\lambda}_-{\partial\over\partial\lambda_-} \right)
\left( \frac{\Lambda}{\gamma} \right) - 
\left(\dot\alpha^*\cdot{\partial\over\partial \alpha^*}+\dot{\lambda}^*_-{\partial\over\partial\lambda^*_-} \right)
\left( \frac{\Lambda}{\gamma} \right) \right) + i \dot{\lambda}_{3R}
\end{eqnarray}
Evaluating the geometry for a particular initial state is now reduced to evaluating the expectation values above.  
This proves to be useful for explicit computation.
For example, explicit computation shows that reference states that are related by the action of a group element lead 
to metrics that can be related by performing a coordinate transformation.
\end{appendix}

\end{document}